\documentclass[twocolumn,english,aps,pra]{revtex4-1}
\usepackage[T1]{fontenc}
\usepackage[latin9]{inputenc}
\usepackage{amsmath}
\usepackage{graphicx}
\usepackage{amssymb}

\makeatletter
\@ifundefined{textcolor}{}
{%
 \definecolor{BLACK}{gray}{0}
 \definecolor{WHITE}{gray}{1}
 \definecolor{RED}{rgb}{1,0,0}
 \definecolor{GREEN}{rgb}{0,1,0}
 \definecolor{BLUE}{rgb}{0,0,1}
 \definecolor{CYAN}{cmyk}{1,0,0,0}
 \definecolor{MAGENTA}{cmyk}{0,1,0,0}
 \definecolor{YELLOW}{cmyk}{0,0,1,0}
 }

\makeatother

\usepackage{babel}

\begin{document}

\title{Vortices in fermion droplets with repulsive dipole-dipole interactions}

\author{G. Eriksson, J. C. Cremon, M. Manninen$^*$ and S. M. Reimann}

\affiliation{Mathematical Physics, LTH, Lund University, SE-22100 Lund, Sweden}

\affiliation{$^*$ Nanoscience Center, Department of Physics, University of
  Jyv\"askyl\"a, FI-40014 Jyv\"askyl\"a, Finland}

\begin{abstract}
Vortices are found in a fermion system with {\it repulsive} 
dipole-dipole interactions, trapped by a rotating quasi-two-dimensional
harmonic oscillator potential. Such systems have much in common with
electrons in quantum dots, where rotation is induced via an
external magnetic field. In contrast to the Coulomb interactions between
electrons, the (externally tunable) anisotropy of the dipole-dipole interaction 
breaks the rotational symmetry of the Hamiltonian. This may cause the otherwise
rotationally symmetric exact wavefunction to reveal its internal  
structure more directly.
\end{abstract}

\maketitle

\section{Introduction}

Rotating quantum fluids have been studied for a long time, with the interest
being spurred by their many fascinating properties. A prominent example are
superfluids such as liquid helium~\cite{donnelly1991}. 
When set rotating in a bucket, for sufficiently large rotation, the quantum
liquid becomes penetrated by quantized vortices,  forming the well-known
Abrikosov lattice. 

After the early prediction of Bose and Einstein in the 1920's~\cite{BEC} 
it was not until 1995 that the condensation of a gas of bosonic atoms into 
a single coherent quantum state could be achieved~\cite{BECexp}. 
Stirring the condensate with lasers or rotating the
trap that confines the dilute atom gas, vortices similar to those in $^4$He
\cite{butts-rokshar-Nature-1999,kavoulakis-et-al-PRA-2000} were observed 
\cite{madison-et-al-PRL-2000,abo2001}
(see also the review by Fetter~\cite{fetter-RMP-2009}). 
Similar states have been realized also
for trapped fermionic atoms with attractive interactions where pairing or
molecule formation can occur (see the reviews~\cite{giorgini2008,bloch2008}). 

Apart from their presence in systems with bosons, vortices have also been predicted to occur 
in fermion systems with purely {\it repulsive} interactions,  such as 
quantum dots -- small man-made electronic systems that can be created in a 
semiconductor heterostructure~\cite{qdotreview}, where the rotation can be induced with an 
external magnetic field. 
For small numbers of confined electrons, it was shown
that these fermionic quantum Hall droplets form vortices in a very similar way
than repulsive bosons set rotating in the 
trap~\cite{saarikoski2004,toreblad-et-al-PRL-2004,viefers-review-2008}. 
This analogy may, in fact, question the commonly accepted view that 
vortices and vortex arrays may be taken as a criterion of  
superfluid properties~\cite{saarikoski-et-al-RMP-2010}. 
Experimentally however, vortices in quantum dots are difficult to detect:
As the electrons are inside a semiconductor crystal, 
probing their properties must typically be done by indirect methods. 
Examples are electron transport 
or magnetization measurements~\cite{saarikoski2005,saarikoski-et-al-RMP-2010}.  
These are, however, strongly hindered by the restricted resolution in the 
conductance spectra, as well as unavoidable sample imperfections.

Here, ultra-cold atomic gases may be the better choice, as they  
typically are very clean, and remarkable characterization techniques have
been demonstrated. For example, it is possible to directly  image
the atomic cloud after expansion. 
Atomic quantum gases usually confine millions of atoms, bringing system sizes
close to the thermodynamic limit.
Serwane {\it et al.}~\cite{serwane2011} however  showed recently that the
confinement of about a dozen of cold fermionic atoms can be reached
experimentally. 

The electrons in a quantum dot interact via long-range repulsive Coulomb 
forces. 
An alternative is provided by atoms or molecules with (either
electric or magnetic) dipole-dipole interactions
providing
a long-range coupling between the particles. 
(There has recently been much interest in 
such dipolar systems, see for example 
Refs.~\cite{koch-et-al-NP-08,lu-et-al-PRL-10,danzl-et-al-NP-10,deiglmayr-et-al-PRL-08,muller-et-al-PRA-11,trefzger-et-al-JPB-11},
and~\cite{lahaye-et-al-RPP-09,baranov-PhysRep-08} for reviews).
In contrast, the very short-ranged van~der~Waals interaction has a limited
effect on spin-polarized fermions due to the Pauli exclusion principle, and might 
not give a similar response to rotation as long-range interactions.

In this paper, we show that vortices and vortex clusters 
occur in a (spin-polarized) fermion droplet with strictly repulsive 
dipole-dipole interactions, in much analogy to
vortices in a quantum dot. While in quantum dots the vortex lattice still 
awaits experimental detection, we suggest that dipolar fermionic atoms with 
repulsive interactions may indeed show a similar vortex lattice than analogous 
systems with bosons. 

\section{Model}

As a simple model (that well resembles that of electrons
in a quantum dot) we consider a few (spin-polarized)
fermions confined in a rotationally symmetric two-dimensional 
harmonic oscillator
in the $xy$--plane, with oscillator frequency $\omega_{0}$. We assume
that the particles are confined in the $z$-direction with a tight
harmonic oscillator, so that the total trapping potential is $V_{\mathrm{trap}}(x,y,z)=\frac{1}{2}m\omega_{0}^{2}(x^{2}+y^{2})+\frac{1}{2}m\omega_{z}^{2}z^{2}$.
The oscillator length $l_{z}$ is here set to be $1/100$
of that in the plane, so that the particles 
occupy the lowest orbital in this direction. 
(In the following, we work in dimensionless 
oscillator units, e.g. lengths are given in units of 
$l_{0}=\sqrt{\hbar/(m\omega_{0})}$,
energies in $\hbar\omega_{0}$ and frequencies in $\omega_{0}$.) 
The rotation is induced
by adding a term $-\Omega \hat L$ to the Hamiltonian, where $\Omega$ is the
rotational (angular) frequency of the rotating trap, and $\hat L$ is the
($z$-projection) of the angular momentum operator.

For the interaction between the fermions we 
assume that their dipole moments are aligned
by an external field. This makes it possible to control the effective
anisotropy of the interaction, breaking the rotational
symmetry in the system. A few other theoretical studies considered 
very similar
setups~\cite{lewenstein-PRL-2007,lewenstein-PRL-2008,qiu-et-al-PRA-2011}. Here,
however, the focus is on the vortex structure.  

For two particles with dipole moments, a general expression for their interaction 
energy is given in e.g. Ref. \cite{lahaye-et-al-RPP-09}. In the present system, 
we assume the dipole moments to be aligned to an axis lying in the $xz$-plane,
such that this axis forms an angle $\Theta$ with the $xy$-plane where the 
particles are effectively confined by the harmonic trapping.
For two point dipoles confined to the $xy$-plane, the interaction
is attractive for $\Theta<\arccos\frac{1}{\sqrt{3}}\approx54.7^{\circ}$.
To avoid the possibility that the system collapses in the case of attractive
interactions, we restrict the
tilting to the interval $54.7^{\circ}\lesssim\Theta\leq90^{\circ}$,
where the interaction is repulsive. 
The coordinate system used here is co-rotating with the trap (at frequency
$\Omega$), meaning that the dipole axis is rotating in the
laboratory frame. The in-plane interaction between the particles is obtained
by analytically integrating their motion along the $z$-direction, where the particles 
are assumed to be in the lowest oscillator state.
This results in the interaction potential \begin{equation}
\begin{aligned}V_{\mathrm{2D}}(r,\phi)= & \frac{D^{2}}{2\sqrt{2}}\frac{e^{\xi/2}}{(l_{z}/l_{0})^{3}}\{(2+2\xi)K_{0}(\xi/2)-2\xi K_{1}(\xi/2)\\
 & +\cos^{2}\Theta[-(3+2\xi)K_{0}(\xi/2)+(1+2\xi)\\
 & \times K_{1}(\xi/2)]+2\cos^{2}\Theta\cos^{2}\phi[-\xi K_{0}(\xi/2)\\
 & +(\xi-1)K_{1}(\xi/2)]\},\end{aligned}
\label{eq:interaction-function}\end{equation}
where $K_0$ and $K_1$ are irregular modified Bessel functions, and  
$\xi=r^{2}/(2(l_{z}/l_{0})^{2})$. The prefactor $D$ is here dimensionless, for an electrical dipole it corresponds to 
$D=\frac{d}{\sqrt{4\pi \epsilon_0}}\frac{\sqrt{m}}{\hbar\sqrt{l_0}}$ where $d$ is the dipole moment, and similarly for a magnetic dipole \cite{cremon-bruun-reimann-2010}.
The expression in Eq.~(\ref{eq:interaction-function}) (or special
cases of it) is given e.g. in Refs. \cite{yi-pu-PRA-2006,komineas-cooper-PRA-2007,lewenstein-PRL-2008,cai-et-al-PRA-2010,cremon-bruun-reimann-2010}.
Only for $\Theta=90^{\circ}$ does the interaction have rotational
symmetry. In this limit, it is qualitatively similar to the electrostatic
Coulomb interaction, however, being more short-ranged. 
For other angles, the two-body term
is spatially anisotropic (cf Fig.~1 in Ref. \cite{cremon-bruun-reimann-2010}).

The ratio $l_z/l_0$ enters above, and as it is typically much smaller than one, at first sight it gives a very large 
interaction strength. However, this ratio also enters elsewhere in the expression so that its final behavior 
is more subtle. It turns out that for long ranges, Eq.~(\ref{eq:interaction-function}) is practically 
independent of $l_z/l_0$ and proportional to $D^2/r^3$. (Long range here corresponds to  
$r\gg l_z/l_0$.) At short range the coefficient $\frac{D^2}{(l_z/l_0)^3}$ does have an impact for the value of the 
interaction potential, but in this study we 
only consider spin-polarized fermions, for which the Pauli exclusion principle eliminates the effect of the 
interaction at short ranges.

The many-body Hamiltonian
\begin{equation}
H=\sum_{i=1}^{N}\biggl(\frac{1}{2}\mathbf{p}_{i}^{2}+\frac{1}{2}\mathbf{r}_{i}^{2}\biggl)+\frac{1}{2}\sum_{i\neq
  j}^{N}V_{\mathrm{2D}}(\mathbf{r}_{i}-\mathbf{r}_{j}) - \Omega \hat L\label{eq:hamiltonian}\end{equation}
is then diagonalized numerically, applying the configuration
interaction method (also called 'exact' diagonalization). 
The commonly used lowest Landau level
approximation~\cite{saarikoski-et-al-RMP-2010} allows us to 
restrict the Hilbert space to a basis of
Slater determinants constructed with the single-particle orbitals
$\psi_{n=0,m\ge0}(r,\varphi)=\frac{1}{\sqrt{m!\pi}}r^{m}e^{im\varphi}e^{-r^2/2}$
of the two-dimensional harmonic oscillator. This is a reasonable approximation
in the limit when the rotational frequency $\Omega$ is close to,
but smaller than, the trapping frequency $\omega_{0}$, and the interaction
is weak. We set the interaction strength $D=0.1$ to assure that
the interaction energy is smaller than the Landau level spacing, which
is of order $\hbar\omega_{0}$. In the
calculations, sufficiently large $m$-values are included so that
the resulting energies are unaffected by this cutoff.

\section{Results}

Characteristic for the response of the system to a monotoneously increasing  
trap rotation $\Omega $ is the change in angular momentum $L(\Omega )$, 
obtained by minimizing the total energy in the
rotating frame of reference as a function of $\Omega $. 
For a superfluid, after the onset of trap rotation, 
the system first remains at rest, $L=0$, until a critical frequency is reached
beyond which vortices begin to penetrate the cloud. 
As $\Omega$ further increases, the angular
momentum changes discontinuously to higher values as
additional vortices appear.  
For a rotating Bose-Einstein condensate, this sequential appearance of
vortices was predicted by applying the mean-field Gross-Pitaevskii 
approach~\cite{butts-rokshar-Nature-1999,kavoulakis-et-al-PRA-2000}, and later
also observed experimentally~\cite{madison-et-al-PRL-2000,abo2001}.  

For electrons in quantum dots, the rotation may be induced by 
an applied magnetic field. In the limit of slow rotation, 
vortices and vortex clusters 
occur even though there is no condensation of bosons present. 
(For a more detailed discussion,
see e.g. Refs. \cite{saarikoski-et-al-RMP-2010,toreblad-et-al-PRL-2004}).

Turning now to fermions with repulsive dipole-dipole interactions, 
for small particle numbers and
relatively slow rotation considered here, we find that indeed, the 
system shows a similar vortex pattern.

Solving the many-body Hamiltonian by direct diagonalization, 
the numerical effort grows considerably with particle number. 
Because of the anisotropic dipole-dipole interaction, angular momentum is not a 
good quantum number, which poses an additional limitation to the system sizes 
that are numerically accessible. For this reason we here limit the study to six
particles confined in the harmonic trap, making use of the Lowest Landau level
appropriate for weak interactions.   

In the case of non-interacting bosons, the many-particle ground state is 
the permanent $|N0000\dots \rangle $ with macroscopic occupancy of the $m=0$
orbital in the LLL basis. For spinless fermions, however, the Pauli principle 
demands single occupancy of the orbitals, leading to the so-called  
``maximum density droplet'' $|111111000\dots \rangle $. This state  
represents the finite-size analog to the Laughlin state for the 
integer quantum Hall effect, and is the fermionic equivalent to the condensate
permanent. The lowest possible angular momentum of 
this Slater determinant is $L_{MDD} = N(N-1)/2$.   

\begin{figure}
\includegraphics[width=1\columnwidth]{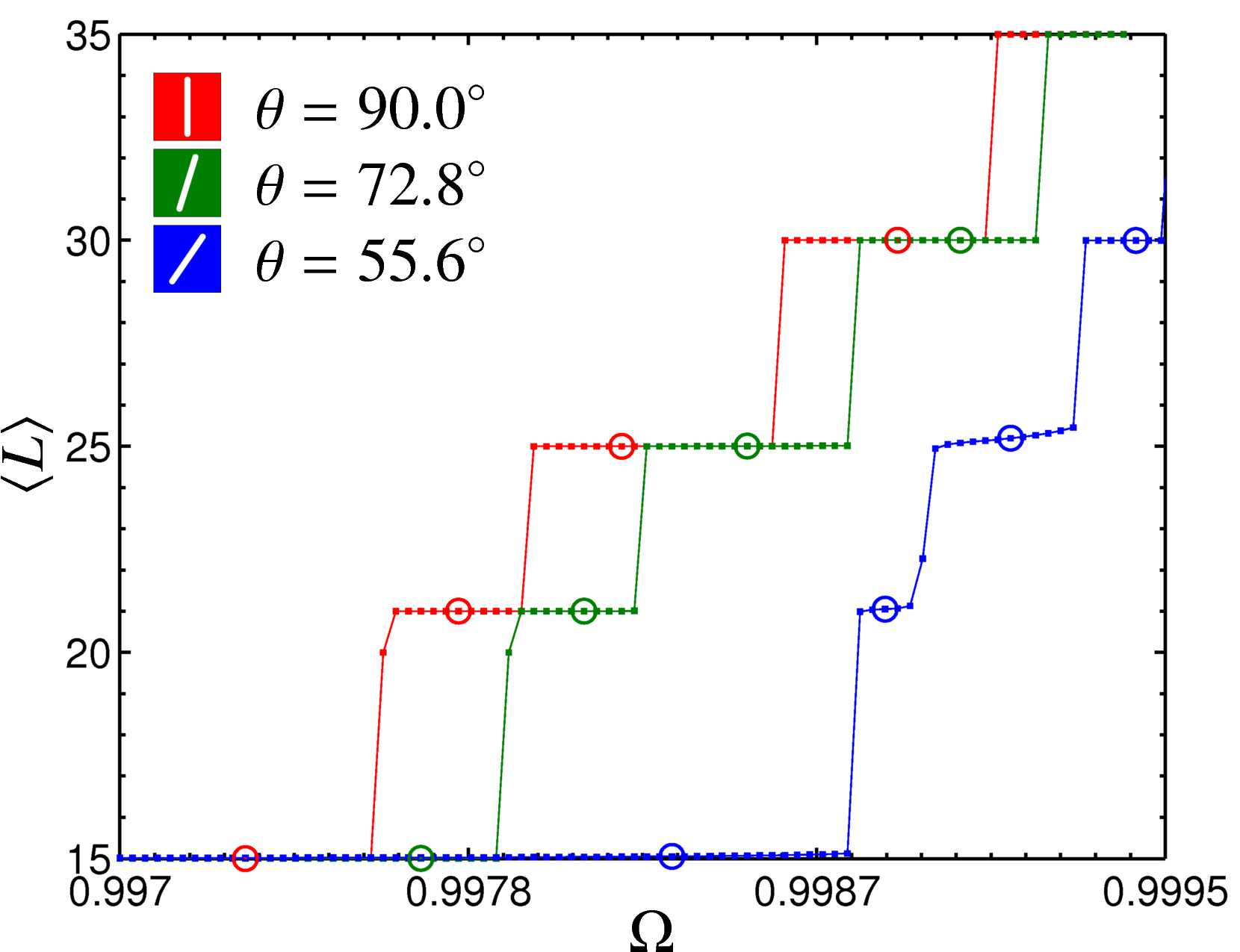}

\caption{\label{fig:Lomega}(Color online) Expectation value of the angular momentum $L$
of the ground state, as a function of the rotational frequency $\Omega $,
for three different tilt angles of the dipoles. The system consists
of $N=6$ (spin-polarized) fermions with dipole-dipole interactions.
As $\Omega $ increases, the angular momentum changes discontinously, 
and vortices penetrate the quantum system. 
The states marked with circles are the ones
which are shown in the following figures.}
\end{figure}

The response of the system with $N=6$ dipolar fermions to the increasing trap 
rotation $\Omega $ 
is shown in Fig.~\ref{fig:Lomega} for different values of the dipolar
tilt angle $\theta $, where the interactions are in the repulsive
regime. (Note that since $L$ is not a good quantum number, we instead calculate the corresponding expectation value that can be non-integer.) Qualitatively,
the resulting picture is very similar as for trapped bosons, or electrons in 
quantum dots.

\subsection{Isotropic interaction}

Let us first consider the case of aligned dipoles with 
$\theta = 90.0^{\circ }$. 
At moderate rotation, the cloud remains at 
rest relative to the angular momentum of the MDD (here, $L_{MDD}=15$), 
until a critical frequency is reached, where the  first step in $L(\Omega )$ 
occurs, and the system aquires angular momentum 
beyond the MDD value. At this frequency a vortex is formed at the center of
the cloud (see discussion below). With increasing rotation, further steps occur 
that are associated with the entry of additional vortices. 
The tilt angle of the dipole axis is found to have little effect on 
$L(\Omega )$: The jumps are mainly shifted to higher values of
$\Omega$ as the angle is lowered. This can be understood by noting that the 
tilt of the dipole angle not only makes the interactions anisotropic, but
also effectively lowers their strength.

The single-particle density distribution, however,  defined
as
$\rho(\mathbf{r})=\langle\Psi|\hat{\Psi}^{\dagger}(\mathbf{r})\hat{\Psi}(\mathbf{r})|\Psi\rangle$, 
can be strongly affected by the tilt, as shown in Fig.~\ref{fig:densities}.
If the dipoles are aligned perpendicular to the plane of motion ($\Theta=90^{\circ}$),
the Hamiltonian has rotational symmetry. Hence, the density of
the associated eigenstates should also be azimuthally symmetric. This
makes it impossible to indentify vortex clusters beyond the simple unit 
vortex (that has azimuthal symmetry). Internally, for the two-vortex solution, the two 
single vortices appear as a pair with two-fold symmetry, while the
three-vortex solution locates the vortices at the corners of an equilateral 
triangle. However, if the Hamiltonian has rotational symmetry these structures 
cannot be visible in the single-particle density. (The pair-correlated 
density, though, can reveal them, as discussed later in the text.) 
This is in contrast with wavefunctions obtained from mean-field approximations 
such as the Hartree-Fock or the Gross-Pitaevskii equations, 
where broken-symmetry solutions are possible.

\subsection{Anisotropic interaction}

\begin{figure}
\includegraphics[width=1\columnwidth]{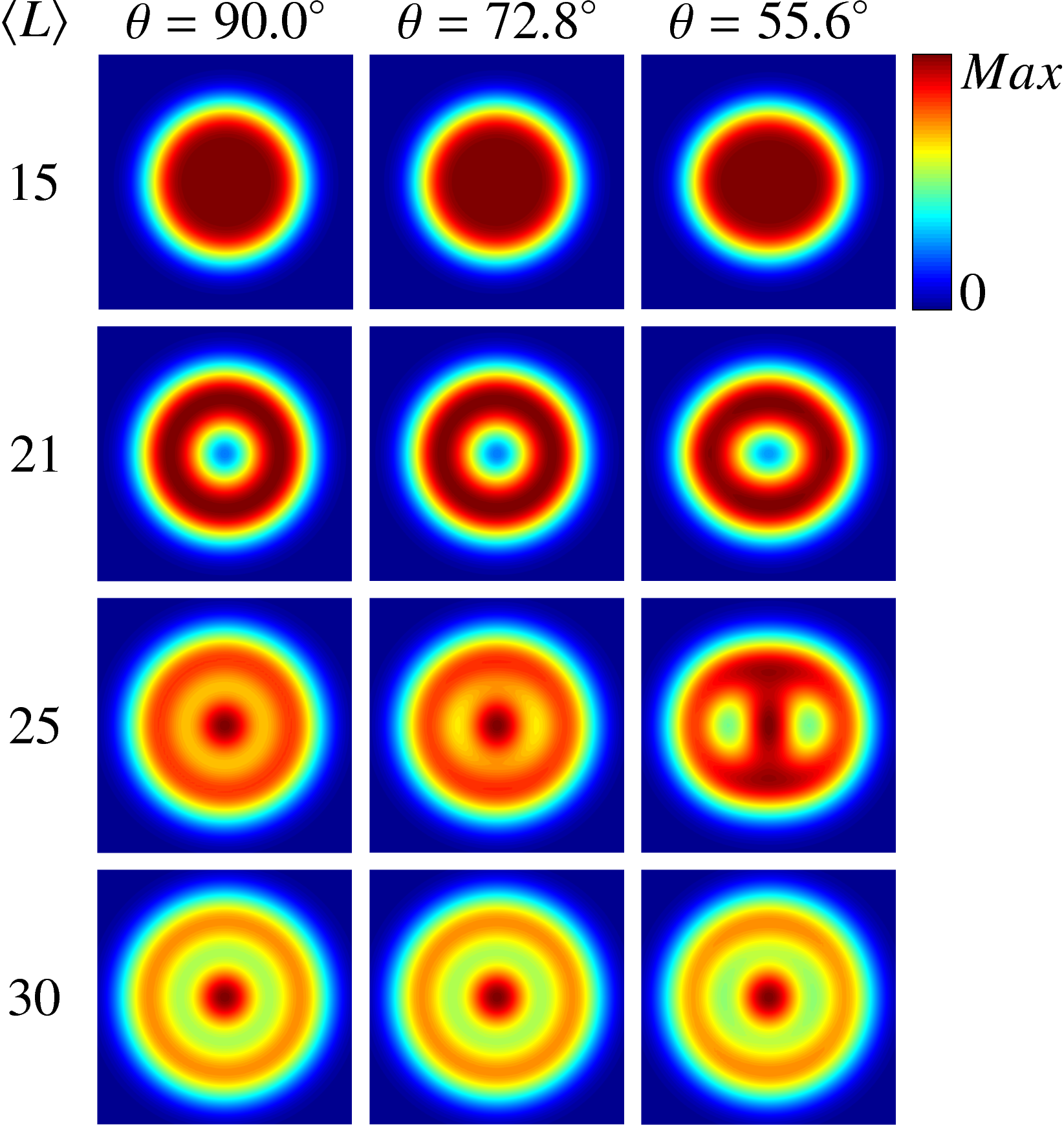}

\caption{\label{fig:densities}(Color online) Probability densities for the states marked
with circles in Fig.~\ref{fig:Lomega}. The listed angular momentum
values are approximate. Each subplot, here and in subsequent figures,
is plotted in the intervals $-4<x<4$ (horizontally) and $-4<y<4$
(vertically). All plots are normalized to have the same peak height, and plotted according to the shown colorbar.}
\end{figure}

When the dipole axis is tilted, the rotational symmetry is broken.
Though some states are not significantly affected by the tilt -- e.g.
the state with a single vortex at the center -- with increasing $\Omega $, 
the state with $\langle L \rangle \approx25$ now shows two clear off-center minima in its density (see Fig. \ref{fig:densities}). 
The overlap
of this state at $\Theta=55.6^{\circ}$ with that at $\Theta=90^{\circ}$
is $93.1\%$, showing that the internal structure of the quantum 
state is largely unchanged despite the seemingly different densities.

The density does not drop to zero at the vortex cores, but this typically 
does not happen in a finite-size system. For example, rotating bosons forming 
a single vortex exhibits zero density in the mean-field solution, but has a 
nonzero density at the vortex core in the exact analytical solution for a 
specific finite number of particles \cite{wilkin-gunn-smith-1998,bertsch-papenbrock-1999,mottelson-1999}.

\subsection{Currents}

\begin{figure}
\includegraphics[width=1\columnwidth]{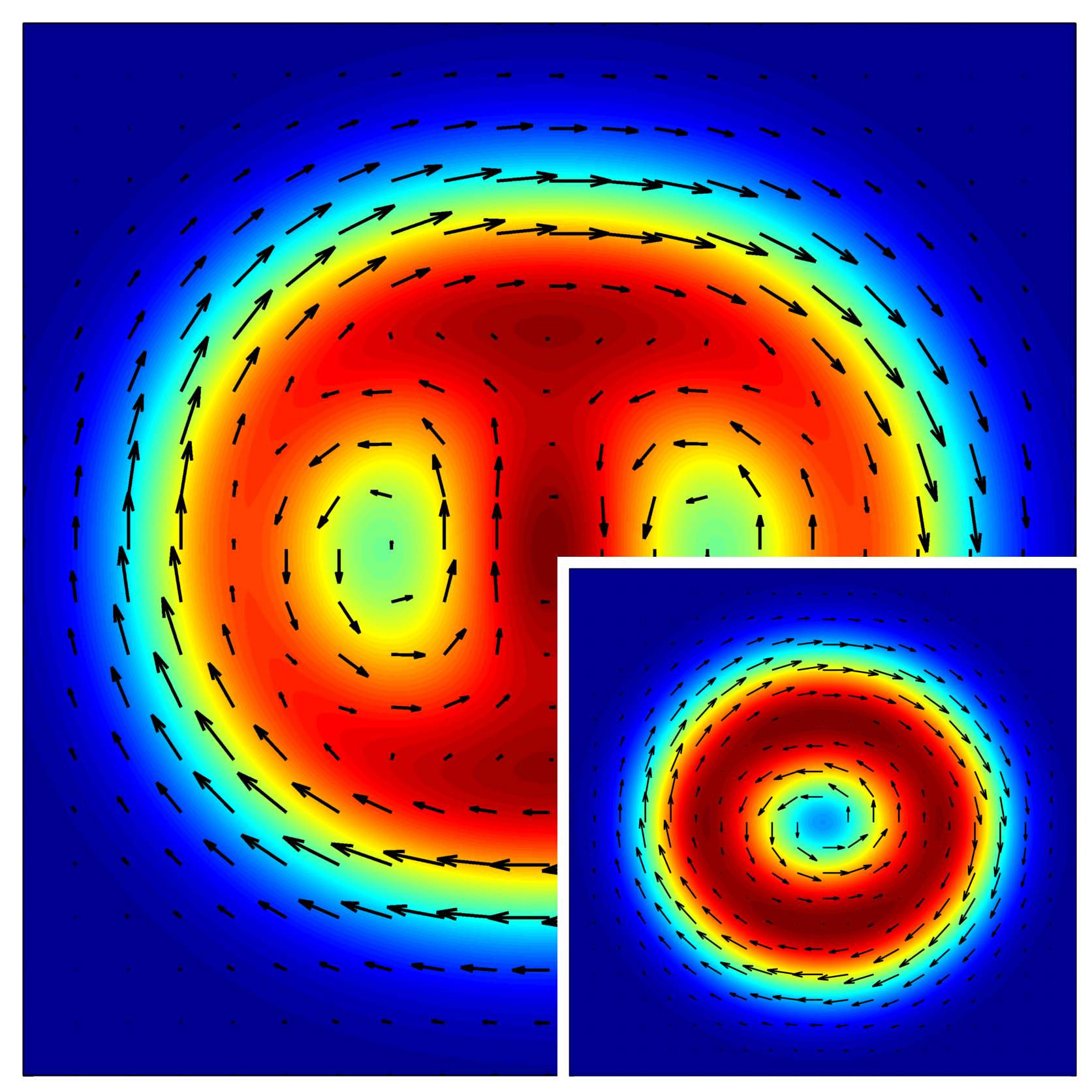}

\caption{\label{fig:currents}(Color online) Probability current (as defined in the text)
for the case $\langle L\rangle\approx25$, when $\Theta=55.6^{\circ}$.
The inset shows the case $\langle L\rangle\approx21$, as comparison.
The black arrows show the current, on top of the single-particle density. 
In both
cases shown, the current can be seen to loop around the density minima.
As an effect of the calculations being performed in the co-rotating
reference frame, the current in the outer parts of the particle cloud
appear to rotate opposite to the current around the vortices. The same plotting conventions as in Fig.~\ref{fig:densities} are
used here.}
\end{figure}

While local minima in the density may indicate vortices, further support
for their existence can be obtained by examining the probability current,
as shown in Fig.~\ref{fig:currents}. The current is here given as
the expectation value of the operator (cf. equations 3--5 in Ref. \cite{saarikoski-et-al-PRB-2005})
\begin{equation}
\hat{\mathbf{j}}(\mathbf{r})=\sum_{i=1}^{N}\frac{-i}{2}[\delta(\mathbf{r}-\mathbf{r}_{i})\nabla_{i}+\nabla_{i}\delta(\mathbf{r}-\mathbf{r}_{i})]-(\Omega\mathbf{e}_{z}\times\mathbf{r})\hat{\rho}(\mathbf{r})\label{eq:current-operator-Jp}\end{equation}
where $\hat{\rho}(\mathbf{r})=\sum_{i=1}^{N}\delta(\mathbf{r}-\mathbf{r}_{i})$,
and $\mathbf{e}_{z}$ is the unit vector of the $z$-direction. 
The last term in the expression originates in the coordinate transformation we use here -- 
the current (as all other calculated quantities) is given in the co-rotating reference frame. 
In principle one could obtain the velocity $\mathbf{v}$ by dividing the current with the density, 
and then the vorticity as $\nabla \times \mathbf{v}$. However, in the regions where the density is very small 
this could yield numerically unstable results, and we here restrict ourselves to 
analyzing the more well-defined current.

For the two-vortex state discussed earlier, Fig.~\ref{fig:currents}
shows the corresponding current which can be seen to circulate around
the density minima, supporting the interpretation of these minima
as vortices. We note that in Fig.~\ref{fig:currents} the inner and
outer regions of the system appear to be rotating in opposite directions
-- this is however an effect of the co-rotating reference frame used
here.

\subsection{Pair-correlated densities}

\begin{figure}
\includegraphics[width=1\columnwidth]{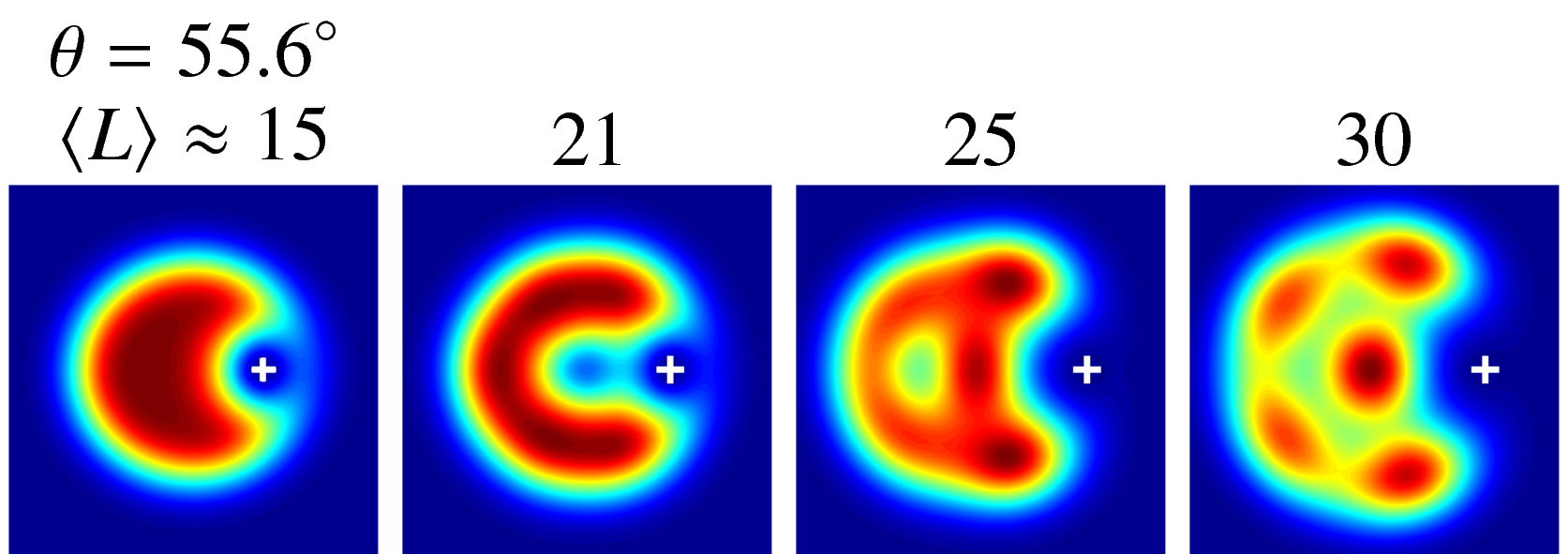}

\caption{\label{fig:paircorr}(Color online) Pair-correlated densities for the states shown
in the right column of Fig.~\ref{fig:densities}, for which the dipole
tilt angle is $\Theta=55.6^{\circ}$. The white cross marks the
position of the reference particle. The same plotting conventions
as in Fig.~\ref{fig:densities} are used here. The positions of the
reference particle are here all on the $x$-axis, and from left to
right $x=$$1.5$, $1.8$, $2.3$ and $2.4$.}
\end{figure}

\begin{figure}
\includegraphics[width=1\columnwidth]{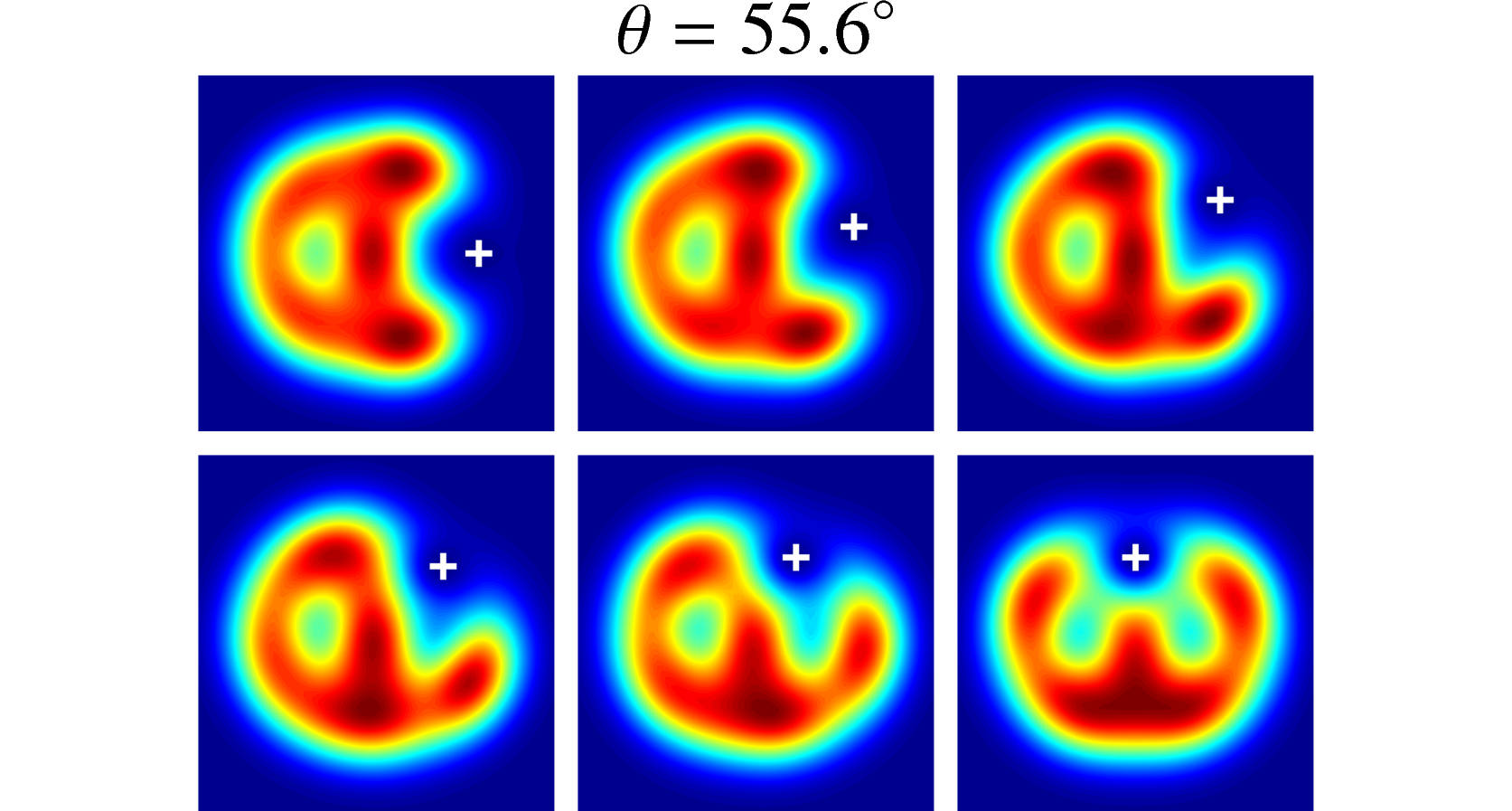}

\caption{\label{fig:paircorr-2vortex}(Color online) Pair-correlated densities for the state
with $\langle L\rangle\approx25$, for $\Theta=55.6^{\circ}$,
with the reference particle placed at different positions (white crosses).
Apart from the necessary fermionic exchange hole, the general structure
of the state is not sensitive to the chosen position. The positions
of the reference particle, starting from top left panel, were $(x,y)=$$(2.4,0)$,
$(2.2,0.6)$, $(1.9,1.2)$, $(1.5,1.5)$, $(0.9,1.7)$ and $(0,1.7)$.
The same plotting conventions as in Fig.~\ref{fig:densities} are
used here.}

\end{figure}

In Fig.~\ref{fig:paircorr} we show pair-correlated densities. Complementing 
the single-particle density, this quantity can give additional information 
about the internal structure of the state. The pair-correlated
density is here defined as $\rho(\mathbf{r},\mathbf{r}')=\langle\Psi|\hat{\Psi}^{\dagger}(\mathbf{r})\hat{\Psi}^{\dagger}(\mathbf{r}')\hat{\Psi}(\mathbf{r}')\hat{\Psi}(\mathbf{r})|\Psi\rangle$,
giving the probability density of simultaneously finding two particles
at positions $\mathbf{r}$ and $\mathbf{r}'$. 
In Fig.~\ref{fig:paircorr-2vortex}
we show pair-correlated densities for the state at $\Theta=55.6^{\circ}$ with 
a reference particle is placed at 
different positions $\mathbf{r}'$ -- showing that the overall structure 
of the density is a stable configuration with two minima, although
somewhat obscured here by the so-called exchange hole around the position 
of the reference particle. (This effect is due to the Pauli exclusion principle, 
which implies that the probability to find any other particle must vanish near 
the position of the reference particle.)

For larger rotation, as $\Omega$ is increased further, the $N=6$
system appears to have too few particles to support more vortices,
and instead it undergoes a transition to a state where the particles
are highly localized, as shown in Fig.~\ref{fig:paircorr}. This
is again very similar to the behavior of electrons in a strong magnetic
field. Though not seen for the parameter regimes considered here,
the anisotropic interaction can be expected to affect the geometries
of such states (see Ref. \cite{qiu-et-al-PRA-2011}).

\section{Summary}

To summarize, we have shown that a quasi-2D trap with 
fermionic particles with dipole-dipole interactions
show a response to induced rotation that is in many ways similar to
that of confined electrons in a magnetic field \cite{saarikoski-et-al-RMP-2010}.
This is perhaps naturally expected given that both the appearing interactions
in both cases are repulsive and long-ranged (as also discussed earlier in 
Refs.~\cite{lewenstein-PRL-2007,lewenstein-PRL-2008}).  
Here, however, we focused on the occurrence of vortices in repulsive 
fermion systems, that may show very similar properties than their 
superfluid cousins with bosons or attractive fermions. 
We found that a system of six dipolar fermions in a harmonic trap 
shows the characteristic step-wise increase of angular momentum as a function 
of the trap rotation. While the cloud first remains at rest, beyond a critical
frequency the angular momentum shows a discontinuous jump by $N$ units, 
and the first vortex 
penetrates the cloud. Additional vortices then occur with increasing $\Omega $. 
We further analyzed how the vortices are affected by a tilt of the dipole
axis. We find that although quantitatively the internal structure of the
quantum state is in fact largely unchanged, the probablity density 
of the two-vortex state is sensitive to the tilt angle, very clearly mapping
out the internal symmetry of the quantum state. 
From a theoretical perspective, this yields a
fortunate situation as it allows the broken
symmetry of the state to be analyzed directly via e.g. the density
and current. For a rotationally symmetric Hamiltonian, this is typically
not possible. In experiments, possibly any disorder or imperfection 
of the trap may already break the symmetry for aligned dipoles at 
$\theta = 90.0^{\circ }$. 

In conclusion, our study suggests that
there may be a possibility to experimentally observe 
vortices in a rotating quantum system without superfluidity, 
composed of 
spin-polarized fermions with repulsive dipole-dipole interactions. 

\begin{acknowledgments}
This work was supported by the Swedish Research Council, and the Nanometer
Structure Consortium at Lund University (nmC@LU).
\end{acknowledgments}


\begin{thebibliography}{25}

\bibitem{donnelly1991} R. Donnelly, {\it Quantized Vortices in Helium II} (Cambridge
University Press) (1991)

\bibitem{BEC} S. Bose, Z. Phys. {\bf 26}, 178 (1924); Einstein, A., 
Sitz.ber. der Preuss. Akad. Wiss., 261 (1924), {\it ibid.}, 3 (1925). 

\bibitem{BECexp} 
M. Anderson, J. Ensher, M. Matthews, C. Wieman, and
S. Cornell, Science {\bf 269}, 198 (1995); K. Davis, M. Mewes, M. Joffe, M. Andrews, and W. Ketterle,  
Phys. Rev. Lett. {\bf 74}, 5202 (1995); 
J.R. Ensher, D. S. Jin, M. R. Matthews, C. E. Wieman, and
E. A. Cornell, Phys. Rev. Lett. {\bf 77}, 4984 (1996); 
E. Cornell and C. Wieman, Rev. Mod. Phys. {\bf 74}, 875 (2002); 
W. Ketterle, Rev. Mod. Phys. {\bf 74}, 1131 (2002).

\bibitem{butts-rokshar-Nature-1999}D. A. Butts and D. S. Rokshar,
Nature \textbf{397}, 329 (1999)

\bibitem{kavoulakis-et-al-PRA-2000}G. M. Kavoulakis, B. Mottelson,
and C. J. Pethick, Phys. Rev. A \textbf{62}, 063605 (2000)

\bibitem{madison-et-al-PRL-2000}K. W. Madison, F. Chevy, W. Wohlleben,
and J. Dalibard, Phys. Rev. Lett. \textbf{85}, 806 (2000)

\bibitem{abo2001} J.R. Abo-Shaer, C. Raman, J. M. Vogels, and W. Ketterle,
Science {\bf 292}, 476 (2001).

\bibitem{fetter-RMP-2009}A. L. Fetter, Rev. Mod. Phys. \textbf{81},
647 (2009)

\bibitem{giorgini2008} S. Giorgini, L. P. Pitaevskii, and S. Stringari, 
  Rev. Mod. Phys. {\bf 80}, 1215 (2008). 

\bibitem{bloch2008} I. Bloch, J. Dalibard, and W. Zwerger, Rev. Mod. Phys.
{\bf 80}, 885 (2008).

\bibitem{saarikoski2004} H. Saarikoski, A. Harju, M.J. Puska, and
  R.M. Nieminen, Phys. Rev. Lett. {\bf 93}, 116802 (2004).

\bibitem{toreblad-et-al-PRL-2004}M. Toreblad, M. Borgh, M. Koskinen,
M. Manninen, and S. M. Reimann, Phys. Rev. Lett. \textbf{93}, 090407
(2004)

\bibitem{viefers-review-2008}S. Viefers, J. Phys.: Condens. Matter {\bf 20}, 123202 (2008)

\bibitem{saarikoski-et-al-RMP-2010}H. Saarikoski, S. M. Reimann,
A. Harju, and M. Manninen, Rev. Mod. Phys. \textbf{82}, 2785 (2010)

\bibitem{saarikoski2005} H. Saarikoski and A. Harju, 
Phys. Rev. Lett. {\bf 94}, 246803 (2005). 

\bibitem{serwane2011} F. Serwane, G. Zurn, T. Lompe, T.B. Ottenstein,
  A.N. Wenz, and S. Jochim, Science {\bf 332}, 6027 (2011). 


\bibitem{koch-et-al-NP-08}T. Koch, T. Lahaye, J. Metz, B. Frölich,
A. Griesmaier, and T. Pfau, Nature Phys. \textbf{4}, 218 (2008)

\bibitem{lu-et-al-PRL-10}M. Lu, S. H. Youn, and B. L. Lev, Phys.
Rev. Lett. \textbf{104}, 063001 (2010)

\bibitem{danzl-et-al-NP-10}J. G. Danzl, M. J. Mark, E. Haller, M.
Gustavsson, R. Hart, J. Aldegunde, J. M. Hutson, and H.-C. N\"agerl,
Nature Phys. \textbf{6}, 265 (2010)

\bibitem{deiglmayr-et-al-PRL-08}J. Deiglmayr, A. Grochola, M. Repp,
K. M\"ortlbauer, C. Gl\"ock, J. Lange, O. Dulieu, R. Wester, and
M. Weidem\"uller, Phys. Rev. Lett. \textbf{101}, 133004 (2008)

\bibitem{muller-et-al-PRA-11}S. M\"uller, J. Billy, E. A. L. Henn,
H. Kadau, A. Griesmaier, M. Jona-Lasinio, L. Santos, and T. Pfau,
Phys. Rev. A \textbf{84}, 053601 (2011)

\bibitem{trefzger-et-al-JPB-11}C. Trefzger, C. Menotti, B. Capogrosso-Sansone,
and M. Lewenstein, J. Phys. B \textbf{44}, 193001 (2011)

\bibitem{lahaye-et-al-RPP-09}T. Lahaye, C. Menotti, L. Santos, M.
Lewenstein, and T. Pfau, Rep. Prog. Phys. \textbf{72}, 126401 (2009)

\bibitem{baranov-PhysRep-08}M. A. Baranov, Phys. Rep. \textbf{464},
71 (2008)

\bibitem{lewenstein-PRL-2007}K. Osterloh, N. Barber\'an, and M.
Lewenstein, Phys. Rev. Lett. \textbf{99}, 160403 (2007)

\bibitem{lewenstein-PRL-2008}M. A. Baranov, H. Fehrmann, and M. Lewenstein,
Phys. Rev. Lett. \textbf{100}, 200402 (2008)

\bibitem{qiu-et-al-PRA-2011}R.-Z. Qiu, S.-P. Kou, Z.-X. Hu, X. Wan,
and S. Yi, Phys. Rev. A. \textbf{83}, 063633 (2011)

\bibitem{yi-pu-PRA-2006}S. Yi and H. Pu, Phys. Rev. A \textbf{73},
061602(R) (2006)

\bibitem{komineas-cooper-PRA-2007}S. Komineas and N. R. Cooper, Phys.
Rev. A \textbf{75}, 023623 (2007)

\bibitem{cai-et-al-PRA-2010}Y. Cai, M. Rosenkranz, Z. Lei, and W.
Bao, Phys. Rev. A \textbf{82}, 043623 (2010)

\bibitem{cremon-bruun-reimann-2010}J. C. Cremon, G. M. Bruun, and
S. M. Reimann, Phys. Rev. Lett. \textbf{105}, 255301 (2010)

\bibitem{saarikoski-et-al-PRB-2005}H. Saarikoski, S. M. Reimann,
E. R\"as\"anen, A. Harju, and M. J. Puska, Phys. Rev. B \textbf{71},
035421 (2005)

\bibitem{qdotreview}S. M. Reimann and M. Manninen,
Rev. Mod. Phys. \textbf{74}, 1283 (2002)

\bibitem{wilkin-gunn-smith-1998}N. K. Wilkin, J. M. F. Gunn, and R. A. Smith, Phys. Rev. Lett. \textbf{80}, 2265 (1998)

\bibitem{bertsch-papenbrock-1999}G. F. Bertsch and T. Papenbrock, Phys. Rev. Lett. \textbf{83}, 5412 (1999)

\bibitem{mottelson-1999}B. Mottelson, Phys. Rev. Lett. \textbf{83}, 2695 (1999)


\end{thebibliography}
\end{document}